\documentclass[amsmath,amssymb,aps,prl,twocolumn]{revtex4-1}
\usepackage{graphicx}
\usepackage{bm}
\usepackage{color}
\usepackage{url}
\usepackage{here}

\begin{document}


\title{Thermal Hall effect in a phonon-glass Ba$_3$CuSb$_2$O$_9$}

\author{K. Sugii}
\email[]{sugii@issp.u-tokyo.ac.jp}
\author{M. Shimozawa}
\author{D. Watanabe}
\author{Y. Suzuki} 
\author{M. Halim}
\author{M. Kimata} 
\author{Y. Matsumoto} 
\author{S. Nakatsuji}
\author{M. Yamashita}
\affiliation{The Institute for Solid State Physics, The University of Tokyo, Kashiwa, 277-8581, Japan}

\date{\today}

\begin{abstract}
A distinct thermal Hall signal is observed in a quantum spin liquid candidate Ba$_3$CuSb$_2$O$_9$. The transverse thermal conductivity shows a power-law temperature dependence below 50 K, where a spin gap opens. We suggest that because of the very low longitudinal thermal conductivity and the thermal Hall signals, a phonon Hall effect is induced by strong phonon scattering of orphan Cu$^{2+}$ spins formed in the random domains of the Cu$^{2+}$-Sb$^{5+}$ dumbbells in Ba$_3$CuSb$_2$O$_9$.
\end{abstract}

\maketitle
\chapter{}
Hall measurements of metals are fundamental tools to investigate their physical properties and play important roles in the study of quantum phenomena, such as the quantum Hall effect or the anomalous Hall effect \cite{Nagaosa}.
On the other hand, no Hall effect is expected in insulators because the Hall effect originates from conduction electrons.
However, a Hall effect of charge-neutral excitations, which is observed as a \textit{thermal} Hall effect (THE), has been predicted to occur in magnetic and non-magnetic insulators \cite{Katsura, Matsumoto, Romhanyi, Lee, Sheng, Wang,Zhang, Qin, Mori,OwerrePRB2016}, thereby providing new insights into the research on quantum spin liquids and other frustrated materials.
So far, three kinds of THEs have been reported in insulators: THEs of magnons in ordered magnets \cite{Onose, Ideue, Hirschberger}, spin excitations in disordered magnets \cite{Hirschberger, Hirschberger2, Watanabe2016}, and phonon THEs \cite{Strohm, Inyushkin}. Among them, the most studied case is the magnon THE which has been observed in ferromagnetic insulators \cite{Onose, Ideue, Hirschberger} and is understood in terms of the Berry phase associated with the magnon bands \cite{Katsura, Matsumoto}. The THEs of spin excitations in paramagnetic states have also been studied theoretically \cite{Katsura, Romhanyi, Lee}; these THEs were reported recently in a spin ice compound Tb$_2$Ti$_2$O$_7$ \cite{Hirschberger2}, a ferromagnetic kagom\'e lattice system \cite{Hirschberger}, and a frustrated kagom\'e lattice system \cite{Watanabe2016}.

In contrast, reports on the THE of phonons have been limited to studies on a dielectric garnet Tb$_3$Ga$_5$O$_{12}$ (TbGG) \cite{Strohm, Inyushkin}. The theoretical origin of the phonon THE has been discussed as a Raman-type interaction between phonons and large spins \cite{Sheng, Wang}, a Berry curvature of phonon bands \cite{Zhang, Qin}, and a resonant skew scattering of phonons by superstoichiometric Tb$^{3+}$ ions \cite{Mori}. Experimentally, it has been determined that there is a large magneto-elastic coupling in TbGG \cite{Araki, Sytcheva}. 
Moreover, the smaller thermal conductivity of TbGG than that of other rare-earth gallium garnets indicates that there is strong scattering of phonons by the Tb$^{3+}$ ions \cite{Inyushkin2010,Slack}. 
These observations imply that the strong phonon scattering is important for the THE. However, the mechanism for the THE in TbGG is poorly understood because the measurement of the THE in TbGG has been limited only to $\sim$5 K \cite{Strohm, Inyushkin}. Furthermore, because TbGG is paramagnetic at 5 K, it is impossible to separate the THE of phonons from that of spins.
Thus, to investigate the phonon THE, it is crucial to observe the THE over a wide temperature range in a single crystal that has strong phonon scattering and no spin THE.

In this Letter, we report the THE in Ba$_3$CuSb$_2$O$_9$ (BCSO) \cite{Nakatsuji, Quilliam, Han, Katayama, Zhou, Do, Ishiguro}, which according to us is an ideal compound to study a phonon THE owing to its strong spin-lattice coupling and the presence of a spin gap.
The hexagonal (6H) perovskite-type BCSO [Fig. 1(a)] is a transparent insulator consisting of a honeycomb-based network of Cu$^{2+}$ ions \cite{Nakatsuji} [Fig. 1(b)], which has been shown to be a short-range random structure \cite{Wakabayashi, Smerald}. 
Stoichiometric single crystals of BCSO keep the hexagonal lattice symmetry without magnetic long-range order down to temperatures much lower than the spin interaction energy ($J/k_{\rm B}\approx$ 50 K) \cite{Nakatsuji, Quilliam}, showing that a quantum spin liquid state is realized in BCSO. In contrast, an off-stoichiometric compound Ba$_3$Cu$_{1-\delta }$Sb$_{2+\delta }$O$_9$ exhibits a Jahn-Teller distortion around 200 K, accompanying a structural transition to an orthorhombic symmetry \cite{Nakatsuji, Katayama} (termed as an ``orthorhombic sample" hereafter).
Multi-frequency ESR measurements have shown that a dynamical Jahn-Teller effect remains only in the hexagonal BCSO even at the lowest temperatures \cite{Han}, suggesting an emergence of an orbital-spin liquid state where the orbital and spin degrees of freedom are entangled with each other \cite{Nasu}.
These results indicate that a strong spin-lattice coupling exists in BCSO, offering a good opportunity to study a phonon THE.
Moreover, a spin gap is formed by a spin singlet formation below $T_g \sim 50$ K as observed in NMR measurements \cite{Quilliam}, enabling us to detect a phonon THE without a spin THE.

\begin{figure}[thb]
\includegraphics[width=0.9\linewidth]{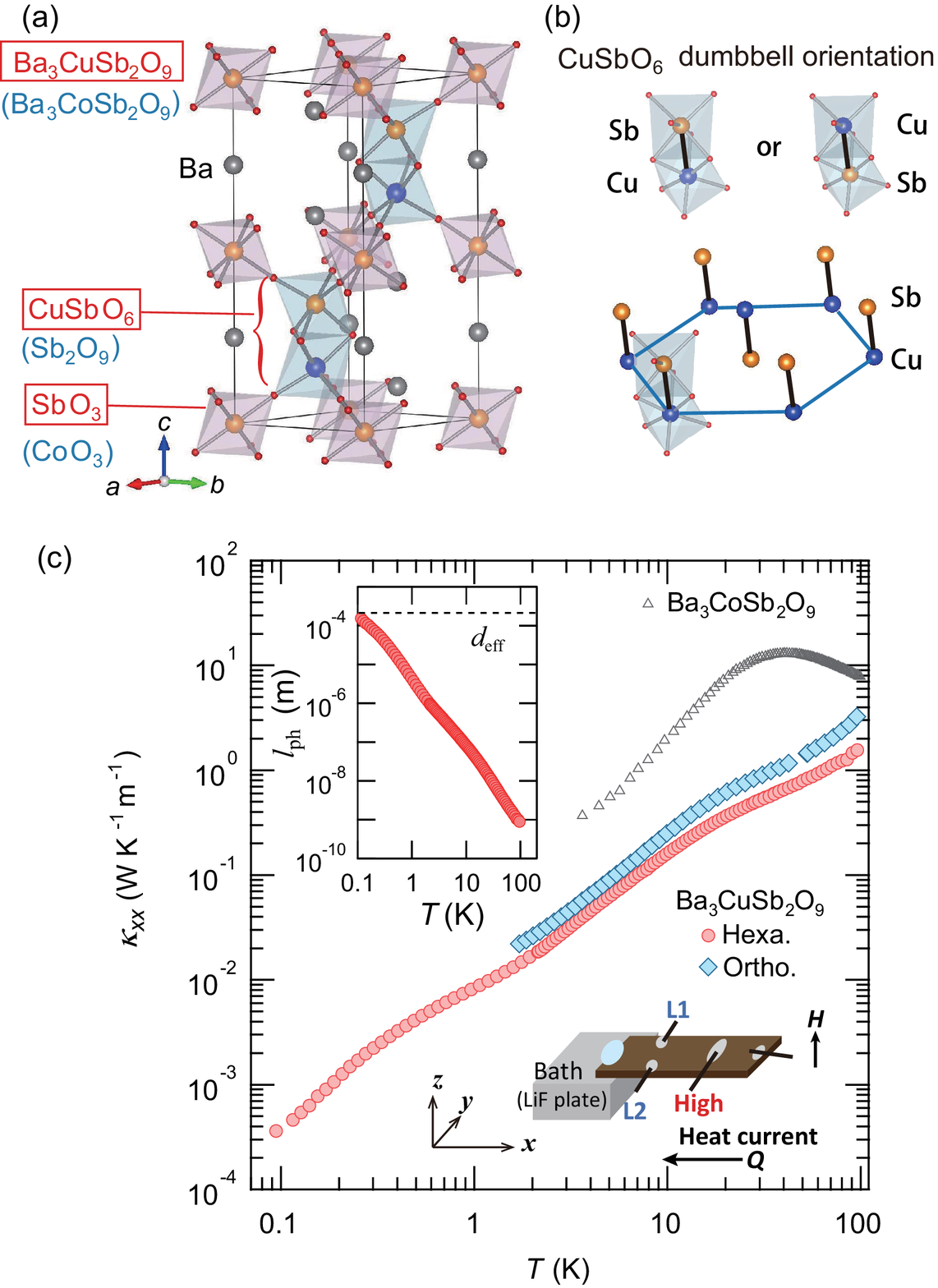}
\caption{(Color online)(a) Crystal structure of Ba$_3$CuSb$_2$O$_9$ \cite{Katayama} and Ba$_3$CoSb$_2$O$_9$ \cite{Doi}. The magnetic Cu$^{2+}$  and Co$^{2+}$ ions are located in different octahedra. 
(b)  Cu$^{2+}$-Sb$^{5+}$ dumbbell structure of Ba$_3$CuSb$_2$O$_9$ and proposed short-range ordered structure of the Cu$^{2+}$-Sb$^{5+}$ dumbbells \cite{Nakatsuji}.
(c) Temperature dependence of the thermal conductivity of hexagonal and orthorhombic Ba$_3$CuSb$_2$O$_9$ and Ba$_3$CoSb$_2$O$_9$. The thermal conductivity data of Ba$_3$CoSb$_2$O$_9$ is taken from Ref.~\cite{Naruse}.
The upper inset shows the temperature dependence of the phonon mean free path of the hexagonal Ba$_3$CuSb$_2$O$_9$.
The lower inset illustrates an experimental setup for the $\kappa_{xx}$ and $\kappa_{xy}$ measurements.}
\label{fig1}
\end{figure}

Longitudinal thermal conductivity ($\kappa_{xx}$) and transverse thermal conductivity ($\kappa_{xy}$) of high-quality single crystals of BCSO were measured in the $ab$ plane in a temperature range of 0.1--100 K. A magnetic field up to 15 T was applied along the $c$ axis and a heat current $Q$ ($\parallel x$) was applied within the $ab$ plane. Three thermometers ($T_{\rm High}, T_{\rm L1}, T_{\rm L2}$) were attached to the sample so that both the longitudinal ($\Delta T_x =T_{\rm High}-T_{\rm L1}$) and the transverse ($\Delta T_y = T_{\rm L1}-T_{\rm L2}$) thermal gradients can be measured at the same setup [see the lower inset of Fig. 1(c) and Supplemental Material for more details]. 

In Fig. 1(c), the temperature dependence of $\kappa_{xx}$ of BCSO is compared with that of an isostructural compound Ba$_3$CoSb$_2$O$_9$ which is a two-dimensional spin $S=1/2$ triangular antiferromagnet with the N{\'e}el temperature of $T_{\rm N} \approx$ 3.8 K \cite{Naruse}.
In magnetic insulators, the thermal conduction is given by the sum of the spin ($\kappa_{xx}^{\rm sp}$) and the phonon ($\kappa_{xx}^{\rm ph}$) contributions.
In Ba$_3$CoSb$_2$O$_9$, it has been shown that $\kappa_{xx}$ in the magnetic $ab$ plane is the almost the same with that perpendicular to the $ab$ plane, suggesting that $\kappa_{xx}^{\rm sp} \approx 0$ and  hence $\kappa_{xx}$ is dominated by $\kappa_{xx}^{\rm ph}$ \cite{Naruse}.
Because hexagonal BCSO and Ba$_3$CoSb$_2$O$_9$ are isostructural, one may suppose that their $\kappa_{xx}^{\rm ph}$ should be almost the same.
However, as shown in Fig. 1(c), $\kappa_{xx}$ of BCSO is about one order of magnitude smaller than that of Ba$_3$CoSb$_2$O$_9$, indicating a strong suppression of $\kappa_{xx}^{\rm ph}$ in BCSO.

The suppression of $\kappa_{xx}^{\rm ph}$ can be clearly seen in the temperature dependence of the mean free path of phonons, $l_{\rm ph}$ [upper inset of Fig. 1(c)]. 
We estimate the upper limit of $l_{\rm ph}$ by $\kappa_{xx} = C_{\rm ph}  v_{\rm ph}  l_{\rm ph}/3$, where $C_{\rm ph}$ is the heat capacity of phonons and $v_{\rm ph}$ is the sound velocity of phonons. Both $C_{\rm ph}$ and $v_{\rm ph}$ are evaluated from the heat capacity data of Ba$_3$ZnSb$_2$O$_9$ \cite{Zhou}. 
Generally, at very low temperatures (typically below $\sim$ 4 K), phonon wave lengths are so long that phonons are not scattered by microscopic defects; only macroscopic objects such as boundaries can scatter phonons, giving rise to a saturation of $l_{\rm ph}$ at low temperatures. 
However, as shown in the upper inset of Fig. 1(c), $l_{\rm ph}$ at 4 K still remains about two orders of magnitude smaller than the effective sample diameter, $d_{\rm eff} = 2\sqrt{ wt/\pi }$. 
We find that $l_{\rm ph}$ reaches $d_{\rm eff}$ only below 0.1 K. 
Such short $l_{\rm ph}$ and the non-saturating temperature dependence even at very low temperatures are characteristic of a glass state in amorphous materials \cite{Berman}, although the hexagonal BCSO is a stoichiometric single crystal \cite{Katayama}.  
A glass-like phonon thermal conductivity, a ``phonon-glass" behavior, has also been observed in crystalline samples, including clathrate compounds \cite{Takabatake}, Tb$_2$Ti$_2$O$_7$ \cite{Li}, and NaCo$_2$O$_4$ \cite{Takahata}. 
In these materials, $\kappa_{xx}^{\rm ph}$ is suppressed by a rattling of the guest atoms \cite{Takabatake}, a strong spin fluctuation \cite{Li}, or a structural disorder by vacancies \cite{Takahata}, resulting in a short $l_{\rm ph}$ with a non-saturating temperature dependence.
The suppressed $\kappa_{xx}$, therefore, indicates that a phonon-glass-like heat conduction is realized in BCSO.

What is the scattering mechanism of the phonons in BCSO?
One may think that the dynamical Jahn-Teller effect suppresses $\kappa_{xx}$. However, as shown in Fig. 1(c), $\kappa_{xx}$ of the orthorhombic sample, where the fluctuation due to the dynamical Jahn-Teller effect ceases below 200 K, is virtually the same as that of the hexagonal sample, which excludes the dynamical Jahn-Teller effect as the origin of the phonon scattering.
Rather, we argue that the strong suppression of $\kappa_{xx}$, which is common to both the hexagonal and the orthorhombic samples, originates from the particular crystal structure of BCSO. 
In contrast to Ba$_3$CoSb$_2$O$_9$, where Co$^{2+}$ ions are located at the center of corner-sharing octahedra \cite{Doi}, the Cu$^{2+}$ ions in BCSO are placed in the CuSbO$_6$ bi-octahedra [see Fig. 1(a)].
The charge imbalance between the Cu$^{2+}$ and Sb$^{5+}$ ions gives rise to the Ising degree of freedom of the  Cu$^{2+}$-Sb$^{5+}$ dumbbells \cite{Nakatsuji}. 
Because the total polarization of the dumbbells should be zero, the Cu$^{2+}$-Sb$^{5+}$ dumbbells tend to form a honeycomb ordered structure as in  Fig. 1(b) \cite{Nakatsuji}. However, a detailed diffuse X-ray study \cite{Wakabayashi} has shown that the honeycomb-like correlation is limited to a short-range scale, and that the dumbbells form a random structure on long length scales. A recent Monte Carlo simulation has also pointed out that the dumbbells form a random domain structure with a broad distribution of domain sizes \cite{Smerald}, which is consistent with the short $l_{\rm ph}$. Therefore, we conclude that a glassy state is formed owing to the broad size distribution of the random Cu$^{2+}$-Sb$^{5+}$ dumbbell structure in BCSO.

The random domain structure also results in a considerable amount of unpaired spins~\cite{Smerald}.
In fact, various measurements \cite{Nakatsuji, Zhou, Quilliam, Do} have shown that there are about 5--16$\%$ of the Cu$^{2+}$ orphan spins in BCSO.
The contribution of the orphan spins to $\kappa_{xx}$ is observed in the field dependence [Fig. 2(a)].
Above 2 K, we find that a magnetic field increases $\kappa_{xx}$ in accordance with a Brillouin curve, which is consistent with a suppression of spin-phonon scattering via the alignment of  free spins under a magnetic field~\cite{Tokiwa}.
On the other hand, $\kappa_{xx}^{\rm sp}$ is decreased by a magnetic field, as observed in a spin-chain compound \cite{Sologubenko2007} and a kagome material \cite{Watanabe2016}.
Therefore, the increase of $\kappa_{xx}$ shows that $\kappa_{xx}^{\rm ph}$ is dominant in BCSO, and that the phonons are scattered by the orphan spins through a spin-phonon coupling in BCSO.

\begin{figure}[bth]
\includegraphics[width=0.9\linewidth]{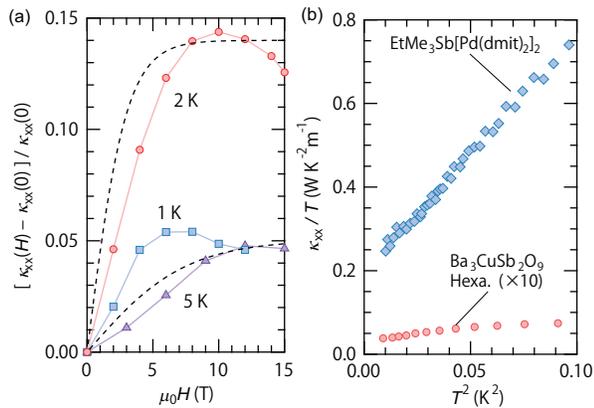}
\caption{(Color online)
(a) Magnetic field dependence of $\kappa_{xx}$ of hexagonal Ba$_3$CuSb$_2$O$_9$ normalized by the zero field values. Dashed lines show Brillouin curves at 2 K and 5 K.
(b) Temperature dependence of $\kappa_{xx}/T$ of the hexagonal Ba$_3$CuSb$_2$O$_9$ at low temperatures.
The data is multiplied by 10 to compare to the data of EtMe$_3$Sb[Pd(dmit)$_2$]$_2$, where a gapless spin excitation is reported \cite{Yamashita}.}
\label{fig2}
\end{figure}

We now discuss the nature of the elementary excitation, which is indispensable to clarify the spin-liquid state in BCSO. So far, NMR measurements \cite{Quilliam} have shown that the intrinsic spin susceptibility drops to zero in the low temperature limit, demonstrating that a spin gap opens below $T_g$. On the other hand, specific heat measurements \cite{Zhou} have reported the presence of a gapless excitation from a residual of the $T$-linear term in the temperature dependence of the heat capacity ($C/T$). 
As shown in Fig. 2(b), we find that the residual of $\kappa_{xx} /T$ as $T\rightarrow 0$ is vanishingly small or zero, demonstrating that there is no gapless excitation in the ground state of BCSO.
This small residual of $\kappa_{xx} /T$ is in stark contrast to that of another QSL candidate EtMe$_3$Sb[Pd(dmit)$_2$]$_2$ [diamonds in Fig. 2(b)] where a gapless spin excitation is reported from the residual of $\kappa_{xx} /T$ \cite{Yamashita}.
The absence of the residual is consistent with the singlet-ground state suggested by the NMR measurements \cite{Quilliam} but is inconsistent with the residual of $C/T$ \cite{Zhou}.
The discrepancy between the residual of $C/T$ and that of $\kappa_{xx}/T$ may indicate that the residual of $C/T$ comes from the localized orphan spins, because $\kappa_{xx}/T$ is sensitive only to itinerant excitations.
We note that a linear temperature dependence of $C$ has been observed in various structural glass materials \cite{Takabatake, Zeller} and spin-glasses \cite{Binder}. Therefore, the linear temperature dependence of $C$ in BCSO may be associated with the phonon-glass behavior observed in $\kappa_{xx}$.

\begin{figure}[tbh]
\includegraphics[width=0.9\linewidth]{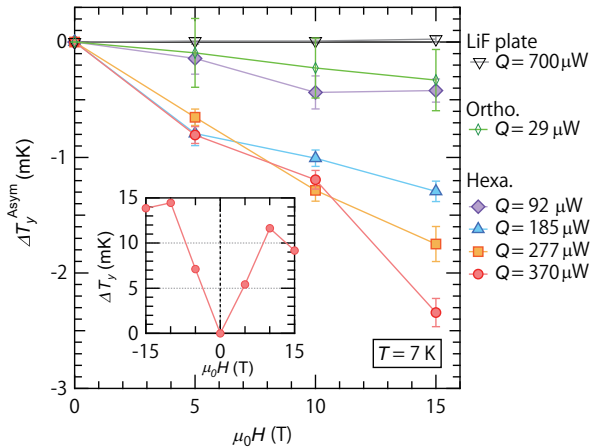}
\caption{(Color online) Transverse temperature difference $\Delta T_y^{\rm Asym}$ at 7 K as a function of the magnetic field and the heat current ($Q$). The data is antisymmetrized with respect to the field direction.
The inset shows the data at $Q =$ 370 $\mu$W before the antisymmetrization.
The data of the orthorhombic sample (open squares) and the background signal from the LiF plate (inverse triangles) are also shown.}
\label{fig3}
\end{figure}

\begin{figure}[tbh]
\includegraphics[width=0.9\linewidth]{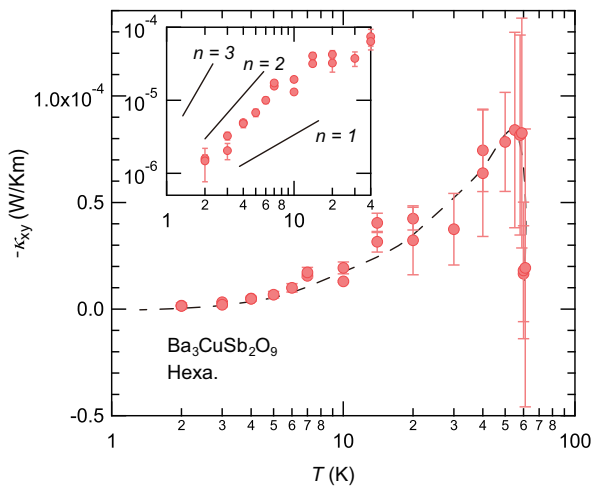}
\caption{(Color online) Thermal Hall conductivity vs temperature of the hexagonal sample. The dashed line is a guide for the eyes. Inset: Log-log plot of the absolute value of the same data below 40 K.
The solid lines show slopes for $\left| \kappa_{xy} \right| \propto T^n (n = 1, 2, 3)$.}
\label{fig4}
\end{figure}

Having established the longitudinal thermal transport in BCSO, we now turn to the transverse heat transport. The inset of Fig. 3 shows the field dependence of $\Delta T_{y} (H)$ at 7 K, where an antisymmetric response is clearly resolved in the hexagonal sample. To cancel the longitudinal response due to a misalignment of the thermal contacts, we antisymmetrized the data with respect to the field direction, $\Delta T_{y}^{\rm Asym} (H)$ =$[\Delta T_{y} (+H)- \Delta T_{y} (-H)]/2$, and plotted the results in Fig. 3. We find that (1) $\Delta T_{y}^{\rm Asym} (H)$ increases almost linearly with respect to the applied magnetic field, (2) $\Delta T_{y}^{\rm Asym} (H)$ increases as the heater power increases [also see the Supplemental Material for the detailed dependence on the heater and the field], and (3) $\Delta T_{y}^{\rm Asym} (H)$ is sufficiently larger than the background signal from the LiF heat bath. These results demonstrate that $\Delta T_{y}^{\rm Asym} (H)$ is not an artifact, but is an intrinsic THE of the sample. 
The THE observed in the transparent insulator immediately means that an unconventional mechanism providing the THE exists in BCSO.

We measured $\Delta T_{y}^{\rm Asym} (H)$ of the hexagonal sample over 2--60 K and evaluated the temperature dependence of $\kappa_{xy}$ at 15 T as shown in Fig. 4. A finite $\Delta T_{y}^{\rm Asym} (H)$ was also observed in the orthorhombic sample (see Fig. 3), but the signal was so small, owing to the small size of the crystal and/or multiple domains caused by the Jahn-Teller distortion, that it is impossible to measure the temperature dependence. As shown in Fig. 4, $\kappa_{xy}$ is negative in the entire temperature range measured, has a peak around 50 K, and decreases at lower temperatures. A log-log plot of the same data below 40 K (inset of Fig. 4) shows that the temperature dependence of $\kappa_{xy}$ asymptotically approaches to $\kappa_{xy} \propto T^{2}$ at low temperature

We suggest that the observed THE is a phonon THE because the spin excitation is gapped \cite{Quilliam} and, hence, $\kappa_{xx}$ is dominated by the phonon transport, which has been confirmed by the field dependence of $\kappa_{xx}$ [Fig. 2(a)] and the vanishing $\kappa_{xx} /T$ as $T\rightarrow 0$ [Fig. 2(b)]. This is in contrast to the cases of the spin THEs where substantial spin thermal conductions are observed \cite{Hirschberger, Hirschberger2, Watanabe2016}.
Moreover, the temperature dependence of the spin thermal Hall conductivity should show an exponential decay below $T_g$ where the spin gap opens, which is in stark contrast to our observation of the power-law temperature dependence of $\kappa_{xy}$. 

In TbGG, it has been pointed out that there are superstoichiometric Tb$^{3+}$ ions in Czochralski-grown crystals, but not in flux-grown ones \cite{Inyushkin2010,Slack}. The THEs are observed only in the Czochralski-grown crystals where $\kappa_{xx}$ is more strongly suppressed than that of flux-grown crystals, which leads the authors in ref. 10 to conclude that the phonon THE is caused by the excess phonon scattering by the superstoichiometric Tb$^{3+}$ ions.
Therefore, it is tempting for us to associate the phonon THE in BCSO with the strong suppression of $\kappa_{xx}$. 
Remarkably, $\kappa_{xy}$ shows a peak just below $T_g$ where $\kappa_{xx}$ also shows a dip followed by a peak at $\sim20$ K [Fig. 1(c)]. This increase of $\kappa_{xx}$ means that spin-phonon scattering by paramagnetic spins is reduced by opening the gap. Given that the orphan spins are formed below $T_g$ and the alignment of the orphan spins determines the field dependence of $\kappa_{xx}$ [Fig.~2(a)], we infer that the phonon THE in BCSO is caused by a spin-phonon coupling between the phonons and the orphan spins. The spin-phonon coupling may also play an important role in realizing the spin-orbital liquid state \cite{Nasu} or the disorder-driven spin-orbital liquid \cite{Smerald} in BCSO. 

The origin of the phonon THE in TbGG has been discussed in several theoretical works.
The mechanisms used in these works can be classified into three groups: the Raman spin-lattice interaction \cite{Sheng, Wang}, the Berry phase of the phonon bands \cite{Zhang, Qin}, and phonon scattering by the 4$f$ Tb$^{3+}$ ions \cite{Mori}.
Both works in the first group show a $T$-linear dependence of $\kappa_{xy}$ \cite{Sheng, Wang}, which is inconsistent with our result. Moreover, the calculation method used in refs. 6 and 7 has been criticized as inappropriate by several authors \cite{Zhang,Qin,Mori}.
The works in the second group predict a $T^n$ dependence ($n=-1$~\cite{Zhang}, 1, or 3~\cite{Qin}), which does not agree with our findings.
These theories take into account the intrinsic mechanism in terms of the Berry phase of the phonon band where thermal transport is ballistic. On the other hand, thermal transport in BCSO is glassy, implying that the dominant mechanism of the THE has an extrinsic origin rather than an intrinsic one.
In the third group, authors in ref. 10 show a nearly $T^3$ dependence of $\kappa_{xy}$, which is different from that of BCSO. 
This discrepancy may be attributed to a difference in scattering mechanisms of phonons by the 4$f$ Tb$^{3+}$ ions with a large quadrupole moment and that of the 3$d$ Cu$^{2+}$ ions.
Thus, all these theories are inconsistent with our results. Further theoretical and experimental research is required to elucidate the origin of the THE and the $T^2$ dependence of $\kappa_{xy}$ in BCSO.
The peak of $\kappa_{xy}$ may further imply that an anharmonic scattering at high temperatures alters the phonon THE or that a spin THE of $\kappa_{xy} > 0$ exists above $T_g$. It also remains, as a future work, to be determined whether other THEs can exist above $T_g$.

To conclude, we have investigated $\kappa_{xx}$ and $\kappa_{xy}$ of Ba$_3$CuSb$_2$O$_9$ (BCSO). We find that $\kappa_{xx}$ is strongly suppressed. The mean free path of phonons remains short even at the lowest temperature, indicating that a ``phonon-glass" like thermal conduction is realized because of the Cu$^{2+}$-Sb$^{5+}$ dumbbell structure of BCSO. The field dependence of $\kappa_{xx}$ shows that the Cu$^{2+}$ orphan spins, formed by the dumbbell domain structure, also scatter phonons. The spin excitation is gapped, which is consistent with the NMR measurements \cite{Quilliam}. We find a finite THE in this transparent insulator and attribute it to a phonon THE. The $\kappa_{xy}$ shows a $T^2$-temperature dependence at low temperature, which should be a key to understanding the THE in BCSO.

\begin{acknowledgments}
We thank H. Katsura, M. Mori, M. Sato, and H. Sawa for valuable discussions, and  K. Torizuka and Y. Uwatoko for technical support. This research was supported by Yamada Science Foundation, Toray Science Foundation, JSPS KAKENHI Grant Numbers 15K17691, 16K17743, 16H02209, Grants-in-Aids for Scientific Research on Innovative Areas (15H05882, 15H05883), and Program for Advancing Strategic International Networks to Accelerate the Circulation of Talented Researchers (No. R2604) from JSPS.
\end{acknowledgments}

\bibliography{reference4}

\clearpage
\noindent
\chapter{\bf \large Supplemental Material for ``Thermal Hall effect in a phonon-glass Ba$_3$CuSb$_2$O$_9$"}

\subsection{\label{sec:level1} Material synthesis and experimental setup}

Two types of  Ba$_3$CuSb$_2$O$_9$ (BCSO) were investigated in our study: stoichiometric hexagonal samples and off-stoichiometric ``orthorhombic" samples.
Both single crystals were synthesized by a flux method described in previous works [25,26]. The detailed difference between two samples are also described in ref. [26].
 
Three Cernox thermometers (CX-1050, Lake Shore Cryotronics, Inc.) were used for the thermal-transport measurements above 2 K and two RuO$_2$ thermometers were used below 2 K.
All thermometers were carefully calibrated in magnetic fields.
We confirmed that the three Cernox thermometers show an identical magnetoresistance which is virtually the same with the results of ref. [S1] within our resolution.
In thermal Hall measurements, three thermometers ($T_{\rm High}, T_{\rm L1}, T_{\rm L2}$) were attached to the sample so that both the longitudinal ($\Delta T_x =T_{\rm High}-T_{\rm L1}$) and the transverse ($\Delta T_y = T_{\rm L1}-T_{\rm L2}$) thermal gradients can be measured at the same setup.
Longitudinal thermal conductivity ($\kappa_{xx}$) and transverse thermal conductivity ($\kappa_{xy}$) are given by
\begin{equation}
\frac{1}{wt} \left( \begin{array}{c} 
      Q \\
      0 \end{array} \right)
= \left( \begin{array}{cc}
      \kappa_{xx} & \kappa_{xy} \\
      -\kappa_{xy} & \kappa_{xx}
    \end{array} \right) 
\left( \begin{array}{c}
      \Delta T_{x}/L  \\
      \Delta T_{y}/w 
    \end{array} \right), 
\end{equation}
where $w$ is the width, $t$ is the thickness, and $L$ is the length between the thermal contacts for $T_{\rm High}$ and $T_{\rm L1}$. 
To avoid background Hall signals from metals around the sample, a LiF single crystal was used as the heat bath.
Furthermore, non-metallic grease was used to attach the sample to the LiF crystal because the silver paste gives rise to a large thermal Hall signal.
We checked that the background Hall signal from the LiF heat bath is small in comparison with the signal from hexagonal BCSO (Fig. 3 in the main text).

\subsection{\label{sec:level2} ESR measurements}

To ensure that the stoichiometric samples keep the hexagonal symmetry at low temperatures, and that the off-stoichiometric orthorhombic samples show the Jahn-Tellar distortion below $\sim$200 K, the ESR measurements were performed for all single crystals used in our thermal-transport measurements by using a conventional X-band ESR spectrometer equipped with a TE$_{\rm 102}$ rectangular cavity. The magnetic field was rotated in the $c$ plane of the single crystal.

In Fig. 5, typical examples of the ESR spectra of hexagonal and orthorhombic samples at 3.6 K are shown.
The ESR signal of the hexagonal BCSO consists of two Lorentzian curves, whereas that of the orthorhombic BCSO consists of four Lorentzian curves.
The $g$-factors of each samples show different angular dependence as shown in Fig. 6.
The $g$-factors of the hexagonal sample show no in-plane angular dependence.
On the other hand, three $g$-factors of the orthorhombic sample show the 180$^{\circ}$ periodicity, and the $g$-factor of the other Lorentzian curve shows no angular dependence ($g \sim 2.08$).
From the comparison with the results in ref. 26, we can conclude that the signal of $g \sim 2.08$ in both samples comes from the Cu$^{2+}$ orphan spins.
The angular dependence of the $g$-factors in the orthorhombic sample can be attributed to three directions of the Jahn-Teller distortion [Fig. 6 (c)].
We confirmed that all the off-stoichiometric orthorhombic samples used in the experiments show the splitting of the ESR signal at low temperatures and the hexagonal ones do not.
Our results are virtually consistent with the previous work [26]. 

\subsection{\label{sec:level2} Magnetic field and heat current dependences of thermal Hall signal}

The linear dependence of the antisymmetrized thermal Hall signal $\Delta T_y^{\rm Asym}$ on the heater power and the magnetic field (Fig. 3 in the main text) can be more clearly seen by plotting  $\Delta T_y^{\rm Asym}/Q$ as a function of the magnetic field (Fig. 7).
The linearity becomes better with increasing the field and/or heater power, and  $\Delta T_y^{\rm Asym}/Q$ converges to a well-reproduced value at the high fields and heater powers.
The deviations from a linear response may be due to our experimental setup and a lack of sensitivity.
The improvement of sensitivity for the Hall measurement at low fields and low heater powers remains as a future work.

\subsubsection{\vspace{-1cm}}
\noindent
{\small [S1] B. L. Brandt {\it et al.}, Rev. Sci. Instrum. {\bf 70}, 104 (1999).}
\subsection{}
\clearpage

\begin{figure}[ht]
\begin{center}
\includegraphics[width= 1\linewidth]{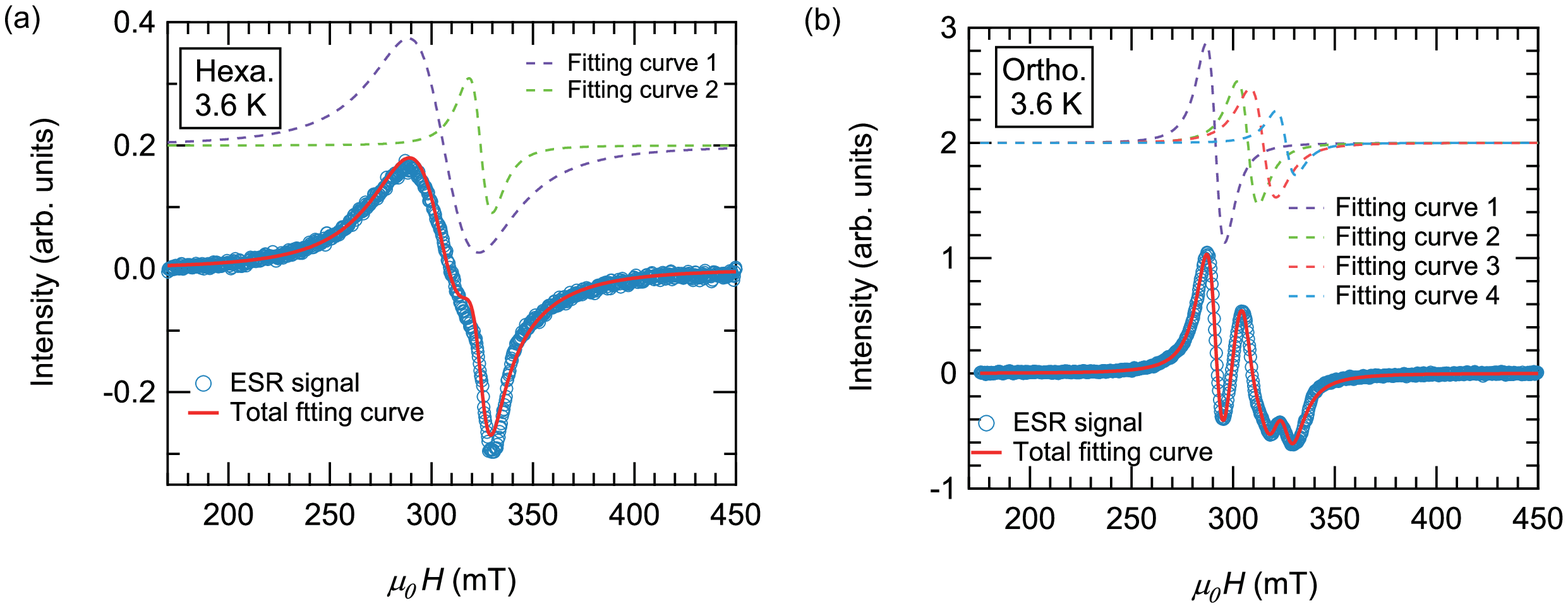}
\caption{ESR signals (blue circle) and Lorentzian fits (dashed curves) for (a) hexagonal and (b) orthorhombic BCSO at 3.6 K. A background signal coming from the cavity is subtracted. Red solid curves are the sum of these Lorentzian fits.}
\end{center}
\label{figs1}
\end{figure}

\begin{figure}[H]
\begin{center}
\includegraphics[width= 0.95\linewidth]{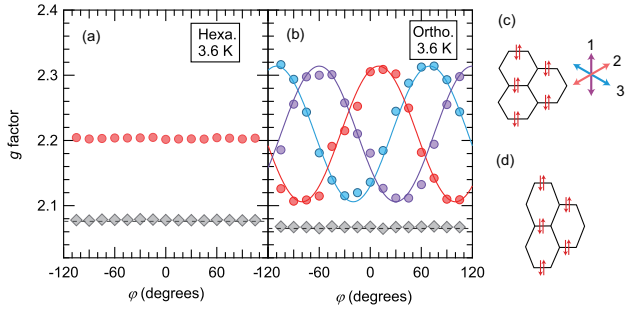}
\caption{In-plane angular dependence of ESR $g$-factors of (a) hexagonal and (b) orthorhombic samples at 3.6 K. The magnetic field was rotated within the $c$ plane. Circle symbols denote the $g$-factors of main signals and diamond symbols denote those of the orphan Cu$^{2+}$ spins. Solid curves and dashed lines are guides for the eye. (c) Schematic illustration of the honeycomb lattice of the Cu spins. The spin singlet is represented by a pair of red arrows. Three arrows (1, 2, 3) denote the three different directions of the Jahn-Teller distortion allowed in the transition from the hexagonal to orthorhombic structure. (d) Schematic illustration of the distorted honeycomb lattice in the orthorhombic sample.
As a consequence of the three domains induced by this static Jahn-Teller distortion, three different angular dependence of $g$-factors are observed in the orthorhombic samples. }
\end{center}
\label{figs2}
\end{figure}

\begin{figure}[H]
\begin{center}
\includegraphics[width= 0.6\linewidth]{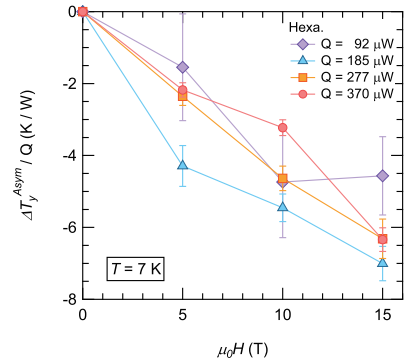}
\caption{Transverse temperature difference divided by heat current at 7 K as a function of the magnetic field. }
\end{center}
\label{figs1}
\end{figure}

\end{document}